# State Complexity of Orthogonal Catenation[*]

Mark Daley[†]    Michael Domaratzki[‡]    Kai Salomaa[§]

October 29, 2018

**Abstract**

A language $L$ is the orthogonal catenation of languages $L_1$ and $L_2$ if every word of $L$ can be written in a unique way as a catenation of a word in $L_1$ and a word in $L_2$. We establish a tight bound for the state complexity of orthogonal catenation of regular languages. The bound is smaller than the bound for arbitrary catenation.

## 1 Introduction

Recently there has been much work on state complexity of operations on regular languages, see for example [4, 8, 14]. Languages $L_1$ and $L_2$ are said to be orthogonal if the catenations of any distinct pairs of words from, respectively, $L_1$ and $L_2$ are distinct (see Section 2 for definitions). Here we consider the state complexity of *orthogonal catenation*, or unambiguous catenation, which is language catenation restricted to pairs of orthogonal languages. Orthogonal catenation has been investigated, for example, by Anselmo and Restivo [1] or the current authors [2]. The current work is motivated by the use of orthogonal language operations in the Code Division Multiple Access (CDMA) multiplexing scheme, used in radio communications.

The state complexity of a related, but essentially different, operation of *unique catenation* was recently investigated by Rampersad et al. [11].

Whereas orthogonal catenation is defined only for pairs of orthogonal languages, the unique catenation of any two languages is defined to consist of those words that can be written in a unique way as a catenation of words from the component languages [11].

We recall the state complexity results for ordinary catenation for general regular languages and some of their restrictions. The definition of a deterministic finite automaton (DFA) can be found in Section 2. The catenation of languages recognized by an $m$-state DFA $A$ and an $n$-state DFA $B$ needs at most

$$m \cdot 2^n - 2^{n-1} \qquad (1)$$

states and there exist worst-case examples where the bound can be reached [9, 10, 12, 13]. The pairs of languages given in [9, 10, 12, 13] that reach this bound are clearly not orthogonal. State complexity of catenation of prefix-free and suffix-free languages is investigated in [5, 7]. Pairs of prefix- and suffix-free languages are necessarily orthogonal. For prefix-free languages state complexity of catenation is significantly less than the bound given by (1), but the situation is not symmetric for suffix-free languages.

Here we investigate state complexity of orthogonal catenation, that is, we consider the situation where $L(A)$ and $L(B)$ are restricted to be orthogonal. The orthogonality condition may appear quite restrictive and, at first sight, one could expect that the state complexity is significantly smaller than in the case of unrestricted catenation. However, we show that the worst-case state complexity of orthogonal catenation is the function (1) divided by two. The bound for orthogonal catenation is approximately two times the worst-case state complexity of catenation for suffix-free languages [7]. We also briefly consider nondeterministic state complexity and transition complexity. In both cases, the condition of orthogonality does not affect the complexity bounds.

We note that the upper bound known for state complexity of unique catenation [11] is larger than (1). The tight bound for the state complexity of unique square of an $n$-state DFA is $n \cdot 3^n - 3^{n-1}$[11]. We can intuitively explain the difference by noting that in the case of unique catenation, it is the automata that are responsible for excluding those words which have multiple factorizations. However, for orthogonal catenation, the languages are known beforehand to obey the appropriate uniqueness restriction. Thus, the bound is considerably lower in this case.

## 2 Preliminaries

In the following $\Sigma$ is a finite alphabet. The set of all words over $\Sigma$ is $\Sigma^*$. For $u, v \in \Sigma^*$, we write $u <_p v$ if $u$ is a proper prefix of $v$.

First we define the orthogonal catenation of languages. The notion can be extended for general binary operations [2]. Let $L$, $L_1$, $L_2$ be languages over $\Sigma$.

**Definition 2.1** *We say that $L$ is the* orthogonal catenation *of $L_1$ and $L_2$, denoted*
$$L = L_1 \odot_\perp L_2,$$
*if the following two conditions hold*

(i) $L = L_1 \cdot L_2$, *and*

(ii) $(\forall u_i, v_i \in L_i, i = 1, 2)$ *if* $(u_1, u_2) \neq (v_1, v_2)$ *then* $u_1 u_2 \neq v_1 v_2$.

Given languages $L_1$ and $L_2$, we define that their orthogonal catenation is undefined if above condition (ii) does not hold. If $L_1 \odot_\perp L_2$ is defined, we say also that the languages $L_1$ and $L_2$ are *(catenation) orthogonal*. Note that in the above statement the order of the languages is significant since the orthogonality relation is not symmetric.

A deterministic finite automaton (DFA) is a five-tuple
$$A = (Q, \Sigma, \delta, q_0, F), \qquad (2)$$

where $\Sigma$ is a finite alphabet, $Q$ is a finite set of states, $q_0 \in Q$ is the start state, $F \subseteq Q$ is the set of accepting states and $\delta : Q \times \Sigma \to Q$ defines the transitions of $A$. In the standard way the transition function $\delta$ is extended to a function $Q \times \Sigma^* \to Q$. Further, for a set $P \subseteq Q$ and $a \in \Sigma$, we use the shorthand $\delta(P, a) = \{\delta(q, a) : q \in P\}$.

We say that a state $q$ is *reachable* from a state $p$ if there exists $w \in \Sigma^*$ such that $\delta(p, w) = q$. A *dead state* is a state $q \in Q - F$ such that only $q$ is reachable from $q$. States $q_1, q_2 \in Q$ are said to be *equivalent* if for any $w \in \Sigma^*$,
$$\delta(q_1, w) \in F \text{ iff } \delta(q_2, w) \in F.$$

A DFA $A = (Q, \Sigma, \delta, q_0, F)$ is a *permutation automaton* if, for each $b \in \Sigma$, the function $\delta(\cdot, b) : Q \to Q$ is a permutation of the set of states $Q$.

The language recognized by a DFA $A$ as in (2) is $L(A) = \{w \in \Sigma^* \mid \delta(q_0, w) \in F\}$. Deterministic finite automata accept exactly the regular languages [13]. Any regular language has a unique DFA with a minimal number of states. In a minimal DFA all states are reachable from the start state and pairwise inequivalent.

For all unexplained notions concerning finite automata we refer the reader to Yu [13].

## 3 Results

First we show that the state complexity of the catenation orthogonal languages recognized, respectively, by an $m$ state DFA $A$ and an $n$ state DFA $B$ can reach $m2^{n-1} - 2^{n-2}$ in cases where $B$ has a dead state. On the other hand, if $B$ does not have a dead state, orthogonality places restrictions on the DFA $A$ that give a corresponding upper bound for the state complexity.

In the following let

$$A = (Q, \Sigma, \delta_A, q_0, F_A), \quad B = (P, \Sigma, \delta_B, p_0, F_B) \qquad (3)$$

be two DFAs. First without making any assumptions on orthogonality, we recall from [13] the construction of a DFA

$$C = (R, \Sigma, \gamma, r_0, F_C) \qquad (4)$$

that recognizes $L(A)L(B)$.

We choose $R = Q \times \mathcal{P}(P) - F_A \times \mathcal{P}(P - \{p_0\})$, $r_0 = (q_0, \emptyset)$, $F_C = \{(q, X) \in R \mid X \cap F_B \neq \emptyset\}$ and the transitions of $\gamma$ are defined by setting for $q \in Q$, $X \subseteq P$, $a \in \Sigma$, $\gamma((q, X), a) = (\delta_A(q, a), Y)$, where

$$Y = \begin{cases} \delta_B(X, a) \cup \{p_0\} & \text{if } \delta_A(q, a) \in F_A, \\ \delta_B(X, a) & \text{if } \delta_A(q, a) \notin F_A. \end{cases} \qquad (5)$$

This construction gives the upper bound (1) for the state complexity of catenation by choosing $A$ to have one accepting state. Also, assuming that $B$ has a dead state $p_{\text{dead}}$, we note that in the DFA $C$ states $(q, X)$ and $(q, X - \{p_{\text{dead}}\})$ are always equivalent. This gives the following upper bound.

**Lemma 3.1** *Let $A$ and $B$ be (minimal) DFAs with $m$ and $n$ states, respectively, and we assume that $B$ has a dead state. Then the state complexity of the language $L(A)L(B)$ is at most*

$$m2^{n-1} - 2^{n-2}. \qquad (6)$$

The next lemma establishes that the upper bound of Lemma 3.1 can be reached by a pair of orthogonal languages.

**Lemma 3.2** *Let $m, n \geq 3$. There exist a DFA $A$ with $m$ states and a DFA $B$ with $n$ states such that $B$ has a dead state and the state complexity of $L(A) \odot_\perp L(B)$ is $m2^{n-1} - 2^{n-2}$.*

**Proof.** Let $\Sigma = \{a, b, c, d\}$ and for $A$ and $B$ we use notations as in (3).

We choose $Q = \{0, 1, \ldots, m-1\}$, $q_0 = 0$, $F_A = \{m-2\}$, and the transitions of $\delta_A$ are defined by setting

1. $\delta_A(0, a) = 0$, $\delta_A(m-2, c) = 0$,

2. $\delta_A(i, b) = i + 1$, $i = 0, 1, \ldots, m-3$,

3. $\delta_A(i, d) = i + 1$, $i = 0, 1, \ldots, m-4$, $\delta_A(m-2, d) = 0$,

4. all transitions not listed in the above cases go to the dead state $m-1$.

The DFA $A$ is depicted in Figure 1.

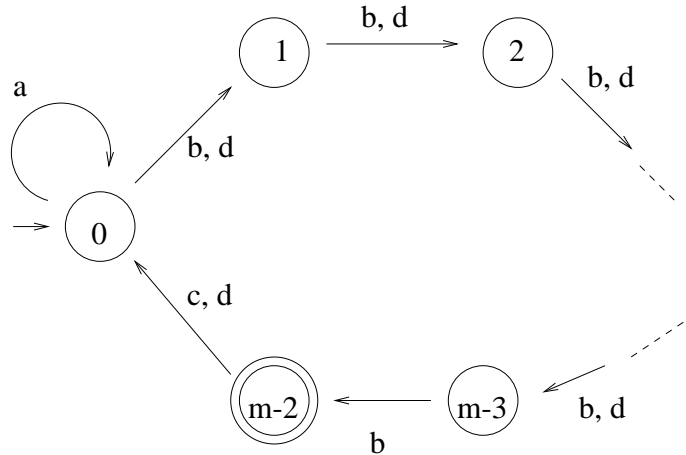

Figure 1: The DFA $A$. The figure does not show the dead state $m-1$ and the transitions into it.

For the DFA $B$ we choose $P = \{0, 1, \ldots, n-1\}$, $p_0 = 0$, $F_B = \{1\}$, and $\delta_B$ is defined by setting

1. $\delta_B(i, a) = i + 1$, $i = 1, \ldots, n-3$, $\delta_B(n-2, a) = 1$,

2. $\delta_B(i, b) = i$, $i = 1, 2, \ldots, n-2$,

3. $\delta_B(i, c) = i$, $i = 2, 3, \ldots, n-2$, $\delta_B(0, c) = 1$

4. $\delta_B(i, d) = i$, $i = 0, 1, \ldots, n-2$,

5. all transitions not listed in the above cases go to the dead state $n-1$.

The DFA $B$ is depicted in Figure 2.

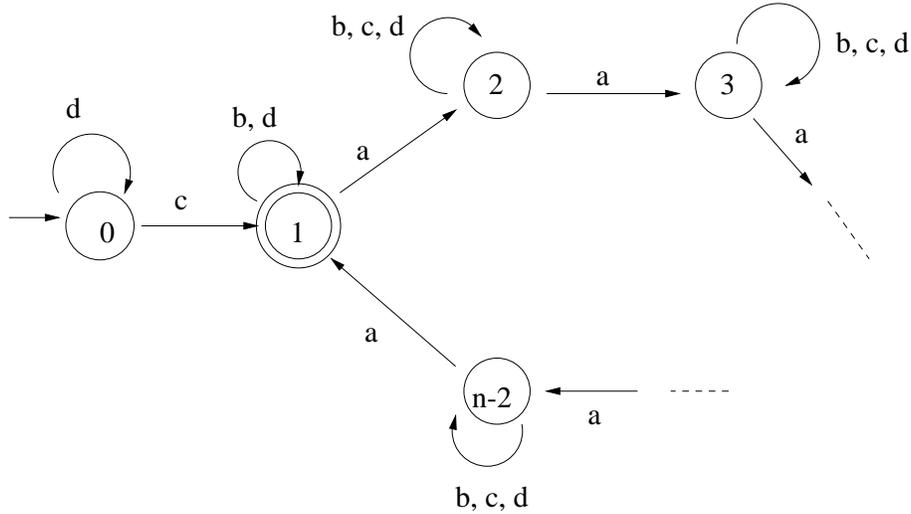

Figure 2: The DFA $B$. The figure does not show the dead state $n-1$.

We show that $L(A)$ and $L(B)$ are catenation orthogonal. For the sake of contradiction, assume that there exist $u_i \in L(A)$, $v_i \in L(B)$, $i = 1, 2$, where

$$u_1 v_1 = u_2 v_2 \text{ and } u_1 <_p u_2. \tag{7}$$

Thus there exists $w \in \Sigma^+$ such that $v_1 = wv_2$.

Since $v_2 \in L(B)$, we can write $v_2 = d^i cz$, $i \geq 0$, $z \in \Sigma^*$ and the number of symbols $a$ in $z$ has to be of the form $j_z \cdot (n-2)$, $j_z \geq 0$. This follows from the observation that in the cycle of $B$ the transitions on symbols other than $a$ are self-loops and the only accepting state is 1.

Now $u_2 = u_1 w$ and $w$ cannot end with a symbol $d$ since $A$ does not accept any words ending with $d$. This means that after reading $w$, the DFA $B$ cannot be in the start state. Hence the computation of $B$ on $v_1 = wd^i cz$ cannot be in state 1 after reading the prefix $wd^i c$ (since the only $c$-transition entering 1 is from the start state). After reading the following $j_z(n-2)$ $a$-transitions the computation cannot end in an accepting state of $B$.

We have seen that (7) produces a contradiction and, consequently, $L(A)$ and $L(B)$ have to be orthogonal.

Since $B$ has a dead state, by Lemma 3.1, we know that the state complexity of $L(A) \odot_\perp L(B)$ is at most $m2^{n-1} - 2^{n-2}$ and hence it is sufficient to show that this value is also a lower bound.

Let $C = (R, \Sigma, \gamma, r_0, F_C)$ be the DFA constructed from $A$ and $B$ as in (4). We denote by $R_1$ the subset of $R$ that consists of all elements $(q, X) \in R$ where $n - 1 \notin X$. Since $R_1$ has $m2^{n-1} - 2^{n-2}$ states, it is sufficient to show that all states of $R_1$ are reachable and pairwise inequivalent in the DFA $C$.

*Claim 1.* All states of $R_1$ are reachable.
*Proof of Claim 1.* First we consider a state

$$(0, X), \text{ where } X \subseteq \{1, \ldots, n-2\}. \tag{8}$$

Using induction on $|X|$ we show that $(0, X)$ is reachable. As the base case, $(0, \emptyset)$ is the start state of $C$. Consider then

$$X = \{j_1, \ldots, j_r\}, \quad 1 \le j_1 < \ldots < j_r \le n-2, \quad r \ge 1. \tag{9}$$

By the silent inductive assumption the state $r_0 = (0, \{j_2 - j_1 + 1, \ldots, j_r - j_1 + 1\})$ is reachable. We note that in the DFA $B$, $b$-transitions on the states in $\{1, \ldots, n-2\}$ are self-loops. Hence

$$\gamma(r_0, b^{m-2}) = (m-2, \{0, j_2 - j_1 + 1, \ldots, j_r - j_1 + 1\}).$$

Then by applying one $c$-transition and shifting the second component by $(j_1 - 1)$ $a$-transitions we get the state $(0, X)$ where $X$ is as in (9).

Next if $Y = \{0\} \cup X$ where $X \subseteq \{1, \ldots n-2\}$ we note that

$$\gamma((0, X), b^{m-2}d) = (0, Y).$$

Finally, from a state $(0, Z)$, $Z \subseteq \{0, 1, \ldots, n-2\}$ we get any state $(i, Z)$, $1 \le i \le m-3$ or $i = m-1$ using only $d$-transitions. Note that $\delta_A(m-3, d)$ is the dead state $m - 1$. Assuming that $0 \in Z$, we have $\gamma((0, Z), d^{m-3}b) = (m-2, Z)$. Recall that $(m-2, Z) \notin R_1$ if $0 \notin Z$.

This concludes the proof of Claim 1.

*Claim 2.* All states of $R_1$ are pairwise inequivalent.
*Proof of Claim 2.* Let $(i_1, X_1)$ and $(i_2, X_2)$ be two distinct states in $R_1$.

First we consider the case where $X_1 \ne X_2$; without loss of generality let $x \in X_1 - X_2$. If $x \in \{1, \ldots, n-2\}$ we note that $\gamma((i_1, X_1), a^{n-1-x}) \in F_C$ since the $a$-transitions in $B$ take $x$ to the accepting state 1. For the same reason $\gamma((i_2, X_2), a^{n-1-x}) \notin F_C$. Note that the $a$-transitions of $A$ keep $i_2$ unchanged or take it to the dead state and hence the $a$-transitions of $C$ do not create any new elements in the second component.

If $x = 0$, then applying the letter $c$ takes $(i_1, X_1)$ to a final state. However, since only the state 0 in $B$ has a transition labelled $c$ which enters the final state 1, $(i_2, X_2)$ is not mapped to a final state by $c$.

Second we consider the case where $X_1 = X_2$ and $0 \leq i_1 < i_2 \leq m - 1$.

(i) First consider the case where $i_1 < m - 2$. Now $\gamma((i_1, X_1), b^{m-2-i_1}c) = (0, Y)$ where $1 \in Y$ and hence $(0, Y) \in F_C$. Note that $m - 2 - i_1 > 0$ and the transitions along $b^{m-2-i_1}$ take $(i_2, X_2)$ to a state $(m-1, Z)$ where $0 \notin Z$. We observe that the last $c$-transition cannot take $(m-1, Z)$ to an accepting state of $C$.

(ii) The only remaining case case is $i_1 = m - 2$, $i_2 = m - 1$. Now in $A$ the word $cb^{m-2}$ takes the state $i_1$ ($= m - 2$) to the accepting state $m - 2$ and hence $\gamma((i_1, X_1), cb^{m-2}c) = (0, Y')$ where $1 \in Y'$. On the other hand, $\gamma((i_2, X_2), cb^{m-2}) = (m-1, Z')$ where $0 \notin Z'$ and hence $\gamma((i_2, X_2), cb^{m-2}c) \notin F_C$.

This concludes the proof of Claim 2 and the proof of Lemma 3.2. ∎

We note that the DFA $B$ used in the proof of Lemma 3.2 has a dead state and, by Lemma 3.1, the result is tight for this type of automata.

Next we consider the situation where $B$ does not have a dead state.

**Lemma 3.3** *Let $A$ and $B$ as in (3) be minimal DFAs such that $L(A)$ and $L(B)$ are catenation orthogonal. If $B$ does not have a dead state, then no accepting state of $A$ can be reachable from itself along a nonempty word.*

**Proof.** For the sake of contradiction assume that there exists $q \in F_A$ where $\delta_A(q, u) = q$, $\delta_A(q_0, w) = q$, for some $w \in \Sigma^*$ and $u \in \Sigma^+$.

Since $B$ does not have a dead state, there exist $0 \leq i < j$ such that $\delta_B(p_0, u^i) = \delta_B(p_0, u^j) = p$ and $\delta_B(p, v) \in F_B$ for some $v \in \Sigma^*$. Thus, the words $w$ and $wu^{j-i}$ are in $L(A)$ and the words $u^j v$ and $u^i v$ are in $L(B)$. Since $w \cdot u^j v = wu^{j-i} \cdot u^i v$ this would imply that $L(A)$ and $L(B)$ are not orthogonal. ∎

Now if $B$ does not have a dead state, and assuming that $L(A)$ and $L(B)$ are orthogonal, we can define an anti-reflexive partial ordering of accepting states of $A$,

$$<_{\text{acc}} \subseteq F_A \times F_A \text{ such that } p_1 <_{\text{acc}} p_2 \text{ if and only if } p_2 \text{ is reachable from } p_1. \tag{10}$$

Furthermore, we know that if $p \in F_A$ is maximal with respect to $<_{\text{acc}}$, the only state reachable from $p$ is the dead state, using also the fact that $A$ is minimal. However, these conditions do not restrict $L(A)$ to be finite.

Under the above conditions, if we construct $C$ as in (4), more than $(m-1)2^{n-1} + 2^{n-2}$ states of $C$ can be reachable only if $A$ has at least $n$ accepting states. To see this, note that if $(q, X)$ is a state of $C$, where $q \in Q$, $X \subseteq P$, and $(q, X)$ is reachable along a word $w$, then the computation of $A$ on $w$ has to enter an accepting state at least $|X|$ times. Since the reachability relation between accepting states of $A$ is an anti-reflexive partial ordering and $A$ has a dead state, it is easy to verify that if $n \geq m$, at least half of the states of $C$ constructed as in (4) must be unreachable.

**Corollary 3.1** *If $n \geq m \geq 3$, the worst-case state complexity of the orthogonal catenation of an $m$-state and an $n$-state DFA is $m2^{n-1} - 2^{n-2}$.*

However, assuming $m > n$, the above observations do not directly prevent the state complexity of orthogonal catenation from exceeding the bound $m2^{n-1} - 2^{n-2}$ in cases where $B$ does not have a dead state. In order to cover also these cases we need to look more carefully at the restrictions that orthogonality places on $A$ in the situation where the second DFA $B$ does not have a dead state.

In the following of this section, unless otherwise mentioned, $A$ and $B$ are always minimal DFAs with notations as in (3), where $A$ has $m$ states and $B$ has $n$ states. Furthermore, we assume that

(A1) $L(A)$ and $L(B)$ are orthogonal, and,

(A2) $B$ does not have a dead state.

In particular, by Lemma 3.3, we know that the accepting states of $A$ can be ordered by a relation $<_{\text{acc}}$ as in (10). Also, $C$ is the DFA constructed from $A$ and $B$ as in (4).

**Lemma 3.4** *Let $p_1, p_2 \in P$, $p_1 \neq p_2$. If*

$$\text{there exists } b \in \Sigma \text{ such that } \delta_B(p_1, b) = \delta_B(p_2, b), \tag{11}$$

*then for any set $\{p_1, p_2\} \subseteq X \subseteq P$ and any $q \in Q$, the state $(q, X)$ cannot be reachable in $C$.*

**Proof.** Recall that the start state of $C$ is $(q_0, \emptyset)$ and new states can be added to the second component according to the rule (5). Thus, if $X$ occurs as a second component of a reachable state there exist $u_1, u_3 \in \Sigma^*$ and $u_2 \in \Sigma^+$ such that $\delta_A(q_0, u_1) \in F_A$, $\delta_A(q_0, u_1 u_2) \in F_A$, $\delta_B(p_0, u_2 u_3) = p_1$ and $\delta_B(p_0, u_3) = p_2$. Here, due to symmetry, we can assume that the predecessor of $p_1$ is first generated by rule (5) after reading $u_1$, and the predecessor of $p_2$ is generated after reading $u_1 u_2$.

Denote $p = \delta_B(p_1, b) = \delta_B(p_2, b)$ where $b$ is as in (11). Since $B$ does not have a dead state, there exists $w \in \Sigma^*$ such that $\delta_B(p, w) \in F_B$.

With the above assumptions we note that $A$ accepts $u_1$ and $u_1 u_2$. On the other hand, $B$ accepts $u_2 u_3 b w$ and $u_3 b w$. This means that the word $u_1 u_2 u_3 b w$ would have two different decompositions as a catenation of words in $L(A)$ and $L(B)$, respectively. ∎

For any distinct states $p_1$ and $p_2$ satisfying the assumptions of Lemma 3.4 we know that $p_1$ and $p_2$ cannot both occur in the second component of a reachable state of $C$. Also, we note that if $q_1 \in F_A$ is a minimal state with respect to $<_{\text{acc}}$, the only reachable state of $C$ with first component $q_1$ is $(q_1, \{p_0\})$, and these observations give the following corollary.

**Corollary 3.2** *Let $A$ and $B$ be DFAs with $m$ and $n$ states, respectively, such that $L(A)$ and $L(B)$ are orthogonal, and, $B$ does not have a dead state. Then the state complexity of $L(A) \odot_\perp L(B)$ can exceed $m 2^{n-1} - 2^{n-2}$ only if the DFA $B$ is a permutation automaton.*

Note that two states as given in the statement of Lemma 3.4 exist if and only if $B$ is not a permutation automaton.

For a state $q \in Q$, by the *valid second components of $q$* we mean the sets $X \subseteq P$ such that $(q, X)$ is a reachable state in $C$.

**Lemma 3.5** *Assume that $B$ is a permutation automaton. Then for every $q \in Q$ such that an accepting state is reachable from $q$ along a nonempty word, there exists $p_q \in P$ such that $p_q$ is not in any valid second component of $q$.*

**Proof.** Let $q \in Q$ be such that

$$\delta_A(q, w) \in F_A \qquad (12)$$

for some $w \in \Sigma^+$.

Since $B$ is a permutation automaton we can choose as $p_q \in Q$ the state with the property

$$\delta_B(p_q, w) = p_0. \qquad (13)$$

Now assume that $p_q$ occurs in some valid second component of $q$. Recalling the rule (5), this means that there exist $u_1, u_2 \in \Sigma^*$ such that $\delta_A(q_0, u_1) \in F_A$, $\delta_A(q_0, u_1 u_2) = q$, and $\delta_B(p_0, u_2) = p_q$.

Choose $v$ to be any nonempty word in $L(B)$. Now using (12) and (13) we note that $A$ accepts the words $u_1$ and $u_1 u_2 w$. On the other hand, $B$ accepts the words $u_2 w v$ and $v$. Since $w \neq \varepsilon$, this produces a contradiction with the orthogonality of $L(A)$ and $L(B)$. ∎

Combining all of the above we can prove the following upper bound.

**Lemma 3.6** *Let $m, n \geq 4$. Let $A$ and $B$ be DFAs with $m$ and $n$ states respectively such that $L(A)$ and $L(B)$ are orthogonal and $B$ does not have a dead state. Then the state complexity of $L(A) \odot_\perp L(B)$ is at most $m2^{n-1} - 2^{n-2}$.*

**Proof.** We give an upper bound estimate for the number of reachable states in $C$. By Corollary 3.2 we can assume that $B$ is a permutation automaton.

Let $q_1, q_2, q_3 \in F_A$ be such that $q_1$ is minimal with respect to $<_{\text{acc}}$ and

$$q_1 <_{\text{acc}} q_2 <_{\text{acc}} q_3.$$

Note that if $<_{\text{acc}}$ does not admit a chain of length three, the valid second components of any state $q \in Q$ would have cardinality at most two, and the claim holds trivially. Without loss of generality we can choose $q_2$ to have distance at most one from any minimal element of $F_A$ in the ordering $<_{\text{acc}}$, and similarly $q_3$ to have distance at most two from any minimal element.

Now the only valid second component of $q_1$ is $\{p_0\}$. Since any computation reaching $q_2$ can have passed at most one accepting state, the possible valid second components of $q_2$ are of the form $\{p_0\} \cup Y$ where $Y$ is a singleton or the empty set. Also since $q_2 <_{\text{acc}} q_3$, by Lemma 3.5, there exists some element of $P$ that cannot occur in $Y$. This means that there exists at most $(n-1)$ possibilities for the valid second component of $q_2$.

By Lemma 3.3, we know that $A$ has a dead state $q_{\text{dead}}$ and an upper bound for the number of valid second components of $q_{\text{dead}}$ is $2^n$.

If $q \in Q - F_A$ is not the dead state, then some accepting state must be reachable from $q$ along a nonempty word. Thus, by Lemma 3.5, an upper bound for the number of valid second components of $q$ is $2^{n-1}$.

If $q' \in F_A - \{q_1, q_2\}$, we know that any valid second components of $q'$ must contain $p_0$ and an upper bound for the number of valid second components is again $2^{n-1}$.

Thus, an upper bound for the number of reachable states of $C$ is

$$1 + n - 1 + 2^n + (m-3) \cdot 2^{n-1}.$$

When $n \geq 4$ the above value is bounded above by the value of $m2^{n-1} - 2^{n-2}$.
∎

Now we can state the main result of this section.

**Theorem 3.1** *For $m \geq 3$ and $n \geq 4$, the worst-case state complexity of the orthogonal catenation of, respectively, an $m$-state and an $n$-state DFA language is $m2^{n-1} - 2^{n-2}$.*

**Proof.** This follows by Lemma 3.1, Lemma 3.2, Corollary 3.1 and Lemma 3.6.
∎

Note that the lower bound construction of Lemma 3.2 assumes $m, n \geq 3$ and the result of Theorem 3.1 does not cover some small values of $m$ and $n$. Also, the construction given in Lemma 3.2 uses an alphabet with 4 letters and the precise state complexity remains open for alphabets of size 2 or 3.

## 4  Nondeterministic State and Transition Complexity

We briefly consider the nondeterministic state complexity of orthogonal catenation. The state complexity of catenation was studied by Holzer and Kutrib [6], who showed that if $L_1$ (resp., $L_2$) has nondeterministic state complexity $n_1$ (resp., $n_2$) then there is an NFA for their catenation with $n_1 + n_2$ states, and this bound is tight. The languages used to show the tightness of this bound are $(a^{n_1})^*$ and $(b^{n_2})^*$, which are obviously orthogonal. Thus, we immediately get the following result:

**Theorem 4.1** *The worst-case nondeterministic state complexity of the orthogonal catenation of, respectively, an $m$-state and an $n$-state NFA language is precisely $m + n$.*

The same witness languages are also used for lower bounds on the transition complexity of regular languages [3], and thus the bounds for transition complexity are also unaffected by orthogonality.

## 5  Conclusions

We have investigated the state complexity of orthogonal catenation. The (deterministic) state complexity of orthogonal catenation is half the state complexity of ordinary catenation. However, the nondeterministic state complexity and transition complexity are unchanged.

The state complexity of orthogonal catenation is distinct from the state complexity of unique catenation, studied by Rampersad *et al.* [11]. It is interesting that the descriptional complexity of these two related catenation operations–both dealing with uniqueness, but in different ways–are markedly different. The situation for nondeterministic state complexity is similar: for orthogonal catenation the nondeterministic state complexity is the sum of the component languages, while in the unique catenation case it is at least exponential [11].

The relation between the two component automata, the presence or absence of dead states, and the role of permutation automata are all factors in the worst case state complexity for orthogonal catenation. The state complexity of the orthogonal square of a regular language is still open, and similar factors will likely play an interesting role in solving this problem.